\documentclass[aip,pof,longbibliography,twocolumn,reprint,a4paper,floatfix]{revtex4-1}

\usepackage{amsmath}
\usepackage{amssymb}
\usepackage{graphicx,color}

\usepackage[T2A]{fontenc}
\usepackage[utf8]{inputenc}

\begin{document}

\title{Linear stability of flow in a 90-degree bend}

\author{Alexander Proskurin}
\affiliation{Altai state technical university
656038, Russian Federation, Barnaul, Lenin prospect, 46
}
\email{k210@list.ru}

\begin{abstract}
The paper considers a two-dimensional flow in a channel, which consists of straight inlet and outlet branches and a circularly 90-degree curved bend. An incompressible viscous fluid flows through the elbow under the action of a constant pressure gradient between the inlet and outlet. Navier-Stokes equations were solved numerically using a high-fidelity spectral/hp element method. In a range of Reynolds numbers, an adaptive selective frequency dumping method was used to get a steady-state flow. It was found that three separation bubbles and vortex shedding can exist in the bend. The modal stability of two- and three-dimensional perturbations was investigated. Critical Reynolds number of the two-dimensional disturbances was found as extrapolation by lower Reynolds number results. It is much greater than three-dimensional one, but the two-dimensional flow could subcritically unstable with respect to the imposed small-amplitude white noise. For three-dimensional perturbations, the dependence of the critical Reynolds numbers on the bending radius is obtained. For a case of a moderate bending radius the neutral curve is provided and eigenfunctions are studied in detail: three-dimensional instability can be caused by periodic or monotonically growing mode, these unstable modes regard to the recirculation bubbles that occur after the bend. 
 
\end{abstract}

% {\bf keywords:} instability, separated flows, channel flow

\maketitle

\section{Introduction}

Complex shape channel flows, when there are recirculation bubbles, flow separation, and periodic motion, are often observed in nature and are widely used in industry. Depending on the Reynolds number, laminar, periodic, turbulent, and locally turbulent patterns can be realized. In the past 30 years, a generation of nonstationary regimes in such flows has been studied using the global stability theory. A review of such research is given in \cite{theofilis:2011:global,taira2020modal}.

An example of flow with separation bubbles is the flow over a backward-facing step. The stability of this flow has been examined extensively and is discussed in several publications \cite{barkley2002three,blackburn2008convective,kaiktsis1996unsteadiness, lanzerstorfer2012global}.  Kaikitis et al. \cite{kaiktsis1996unsteadiness} investigated two-dimensional stability and found that the flow was globally-stable according to the linear analysis if $Re \leq 1000$. They show, that a large part of the flow domain is convectively unstable to the sustained upstream-generated finite-amplitude disturbances for Reynolds numbers $700\leq Re \leq 2500$. Also, Kaikitis et al. assumed that the computations may perform global unsteadiness due to discretization errors that mask the convective instability of the flow. 

Barkley et al. \cite{barkley2002three} have presented the results of a study of the two- and three-dimensional linear stability of this flow. They  found the critical Reynolds number for the three-dimensional perturbations($Re_{\ast}=748$) and have shown that the lower boundary of the critical Reynolds number for two-dimensional perturbations is much higher($Re_{\ast 2D}>1500$). It proves the leading role of the three-dimensional perturbations. Barkley et al. also explain that the instability, according to the linear theory, is associated with the first recirculation bubble, which occurs immediately after the step, and the centrifugal instability is responsible for generating secondary flow with the separation zone.

Lanzerstorfer et al. \cite{lanzerstorfer2012global} considered the backwaed-facing step flow stability for a systematic variation of the step height. It was found that the base flow lost its stability with respect to three-dimensional modes of different types, depending on the expansion ratio. Physical nature of instabilities is investigated by an energy transfer analysis. For very large step height, the instability has a centrifugal nature. In the case of moderate high step, an elliptical mechanism of perturbation growth is the most intense. For small-height steps, the base flow loses stability under the influence of lift-up mechanism. For large step height, the major mode is periodic, and when the height decreases, the instability begin to be determined by a monotonic mode.

Blackburn et al. \cite{blackburn2008convective} examined optimal disturbances growth. They found that three-dimensional optimal disturbances had slightly larger growth in comparison to the two-dimensional. The three-dimensional mode appears near the step and gains energy downstream by the inviscid Orr mechanism and interaction of the Kelvin–Helmholtz instabilities of the two separated bubbles. Also, Blackburn et al. investigated the three-dimensional motion arising under the action of the random inflow noise. They observed a flow with a strong dominance of the two-dimensional dynamics.

Griffith et al. \cite{griffith2007wake} studied the flow in a partially blocked channel. On one side of the channel there was a semicircular bump. The flow was investigated for range of the bump heights, the formation of the main recirculation bubble, minor bubbles, and vortex shedding at higher Reynolds numbers were found. The stability of this flow with respect to three-dimensional perturbations was studied. It was discovered that the critical Reynolds number decreases for higher blockage ratios and that the instability of this flow is determined by the elliptical mechanism rather than the centrifugal one.

Zhang\&Poth\'erat \cite{zhang2013influence} investigated flow in a 180-degree sharp bend. In the output branch of the bend, as the Reynolds number increases, the following patterns can exist: laminar, the appearance of the first recirculation bubble near the inner wall, and the appearance of the second recirculation bubble near the outer wall downstream of the first bubble. If the Reynolds number increases further, vortex shedding can occur in the outlet branch. It was also established that the flow patterns are similar to the flow near a circular cylinder, except for the symmetry, and that the two-dimensional dynamics determine the main features of the three-dimensional flow. Sapardi et al. \cite{sapardi2017linear} studied the stability of this flow. It was shown that the flow can be unstable concerning to the three-dimensional periodic perturbations. Unstable modes are associated with the first recirculation bubble. A two-dimensional nonlinear stability analysis was also made and the hysteresis of the critical Reynolds numbers was found.

Heskestad\cite{heskestad1971two} studied the air flow in a sharply curved channel experimentally and compared the results with the predictions free-strealines theory. Yamashita et al. \cite{yamashita1986fluid} investigated the flow of water and air in a two-dimensional sharp 90-degree bend. They obtained data on the pressure drop and the distribution of the average velocity, velocity fluctuations, the streamlines and the energy spectrum of velocity fluctuations, detected laminar and turbulent types of the flow, observed experimentally a vortex shedding in the range $Re=172.5-750$ (Reynolds number converted to this article parameters) where the first number was the lower boundary for the unsteady motion.

Matsumoto et al.\cite{matsumoto2016two} studied systematically two-dimensional flow patterns in a sharply bent channel for angles less than 90 degrees and describe recirculation bubbles in the outer angle of the bend and near the inner wall of the outlet branch right after the bend. They also observed the periodic vortex emission in the outlet branch and have described conditions for each pattern. Matsumoto et al. \cite{matsumoto2016two} especially notes that the intermediate regimes between the turbulent and the laminar in the bend even are poorly studied. 

Hurd\&Peters\cite{hurd1970analysis}, Orlandi\&Gunsolo\cite{orlandi1979two} studied numerically the laminar flow in a smoothly curved channel and Hurd\&Peters\cite{hurd1970analysis} compared the calculated velocity profiles with their experimental results. Kotb et al. \cite{kotb1988numerical} presented results of a numerical study of the flow in a curved channel were taking attention to the size of the separation regions dependence on the Reynolds number and the bending radius.

Donghun\&Seung Park\cite{park2014streamwise} studied nonlinear stability of the smooth bent flow in the case of small bent angles which were less than 60 degrees and a large bending radius. For such small curvature stability modes shape close to the case of the straight Pouiseuille flow. They describe the nonlinear interaction of the oblique waves, and the amplification of streamwise streaks and vortices. The amplitude of the streaks increases in the bend and the outlet branch, but the streamwise vortices increases only in the bend.

Niu\&Dou\cite{niu2013stability}, Lupi et al.\cite{lupi2020global} studied the stability of the flow in 90-degree pipe bends. Niu\&Dou considered the square cross-section duct using energy gradient theory. Lupi et al.\cite{lupi2020global} investigated a bend of circular pipe and they found using direct numerical modeling that the flow was stable for $Re<2500$. A pair of counter-rotating Dean vortices was observed. The presence of two recirculation regions is detected inside the bend: one on the outer wall and the other on the inner side right after the bend. At $Re>2500$, a periodically oscillating flow was caused by a linear mode, which becomes growing at $Re=2531$. Lupi et al.  concluded that the disturbances increase inside the outer recirculation bubble due to a large shear in the backflow.

The problem is defined in section \ref{a30_problemFormulation}. The numerical method presentation and calculations set up  are described in section \ref{a30_methodology}. Sections \ref{a30_flowRegimes}, \ref{a30_2DStabNonlin} outline possible flow regimes and sections \ref{a30_2DStab},\ref{a30_3DStab} describes results of the investigations of two- and three-dimensional stability. The conclusion \ref{a30_conclusion} summarizes the main findings.

\section{Problem formulation}\label{a30_problemFormulation}
\begin{figure}[t]
\begin{center}
\includegraphics[width=0.5\textwidth]{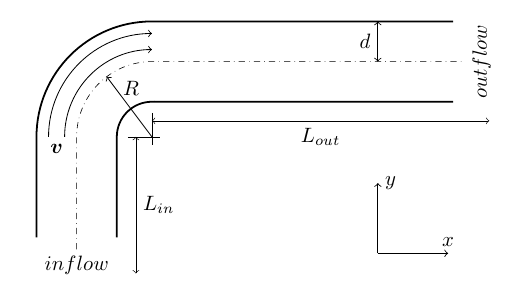}
\end{center}
\caption{The bent channel}\label{a30_LShapeGeometry}
\end{figure}
Figure \ref{a30_LShapeGeometry} shows a bent flow. The channel consists of two parallel impermeable surfaces.  The length of the straight inlet and outlet sections is equal $L_{in}$ and $L_{out}$ respectively. The bending radius on the centre line is equal to $R$. The distance between the surfaces is constant and is equal to $2d$.  A bending parameter is $\delta=R/2d$. The paper considers the case of $\delta=1$, $L_{in}=10$ and $L_{out}=60$. The incompressible viscous fluid flows under the action of a constant pressure difference between the inflow and the outflow. The Reynolds number $Re = \frac{Ud}{\nu}$ was introduced, where $U$ is the maximum velocity of the laminar Pouiseuille flow, and $\nu$ is the fluid viscosity. The Navier-Stokes equation is
\begin{equation}
\label{a30_NSEquation}
\begin{aligned}
\frac{\partial \boldsymbol{V}}{\partial t} + \left(\boldsymbol{V}\cdot\nabla \right)&\boldsymbol{V} = -\nabla p +\frac{1}{Re}\Delta \boldsymbol{V}+\boldsymbol{F},\\
\nabla \cdot &\boldsymbol{V} = 0,
\end{aligned}
\end{equation}
where $\boldsymbol{V}$ is the velocity, $\boldsymbol{F}$ is the external force, and $p$ is the pressure. At the channel walls, the no-slip condition $\boldsymbol{V}=0$ holds true. The Pouiseuille parabolic profile $V_y=1-x^2$ is imposed at the inflow. At the outflow, the standard boundary conditions is imposed\cite{barkley2002three, sapardi2017linear,lanzerstorfer2012global,griffith2007wake}
\begin{equation}
\frac{\partial \boldsymbol{V}}{\partial x} = 0, \quad p=0.
\end{equation}
On all other boundaries the pressure satisfies the high-order Newmann condition\cite{karniadakis1991high}.

If $\boldsymbol{U}(x,y)$ is a steady-state solution of (\ref{a30_NSEquation}), this equation can take a linearized form
\begin{equation}
\label{a30_NSLinEquation}
\begin{aligned}
\frac{\partial \boldsymbol{v}}{\partial t} + \left(\boldsymbol{U}\cdot\nabla \right)&\boldsymbol{v}+\left(\boldsymbol{v}\cdot\nabla \right)\boldsymbol{U} = -\nabla p +\frac{1}{Re}\Delta \boldsymbol{v},\\
\nabla \cdot &\boldsymbol{V} = 0,
\end{aligned}
\end{equation}
where $\boldsymbol{v}(x,y,z,t)$ and $p(x,y,z,t)$ are the small disturbances. Since the baseflow $\boldsymbol{U}(x,y)$ is two-dimensional, the disturbances may be written as
\begin{equation}
\label{a30_Disturbance}
\begin{aligned}
\boldsymbol{v}(x,y,z,t) &= \boldsymbol{v}(x,y)e^{(\sigma+i\omega) t+i\beta z}+\text{c.c.},\\
p(x,y,z,t) &= p(x,y)e^{(\sigma+i\omega) t+i\beta z}+\text{c.c.},
\end{aligned}
\end{equation}
where $\sigma$ is the disturbance growth-rate, $\omega$ is the disturbance frequency, and the spanwise wave number is $\beta = \frac{2\pi}{\lambda}$, where $\lambda$ is the spanwise wavelength. The substitution (\ref{a30_Disturbance}) allows  to decouple the three-dimensional stability problem (\ref{a30_NSLinEquation}) to a series of two-dimensional problems for each value of $\beta$, which varies continuously. This analytical transformation was used by \cite{theofilis:2011:global,barkley2002three,sapardi2017linear} and many others.

Boundary conditions for the disturbance have the form
\begin{equation}
\begin{aligned}
&\boldsymbol{v} = 0 \; \text{at the walls and the inflow},\\
&\frac{\partial \boldsymbol{v}}{\partial x} = 0, \; p=0 \; \text{at the outflow}
\end{aligned}
\end{equation}

\section{Methodology, numerical method and convergence}\label{a30_methodology}

The baseflow and stability calculations were made by the open source spectral/hp element framework Nektar++ \cite{moxey2020nektar++}. The multi-element formulation gives geometric flexibility by comparison to the single-domain spectral methods, and allows to implement the high-order approach to complex shape flows. High-order methods can improve the reliability, accuracy, and computational efficiency of calculations. 

To integrate the equations (\ref{a30_NSEquation}), Nektar++ uses the well-known splitting scheme that decouples the velocity and the pressure\cite{karniadakis1991high}. From equations (\ref{a30_NSLinEquation}) a linear operator $\boldsymbol{A}$ can be constructed:
\begin{equation}
\label{article30.stab_eigproblem}
\boldsymbol{v}(x,y,z,\tau) = \boldsymbol{A}(\tau)\boldsymbol{v}(x,y,z,0) = \lambda(\tau) \boldsymbol{v}(x,y,z,0),
\end{equation}
where $\tau$ is a time interval. $\boldsymbol{A}(\tau)$ is constructed numerically by the splitting procedure in the same way as in the case of  nonlinear equations. In order to find an eigenvalue $\lambda(\tau)$, it is convenient to construct a Krylov subspace
\begin{equation}
\label{article30.stab_krylovsbspace}
K_n(\boldsymbol{A},\boldsymbol{v}_0) = span\{\boldsymbol{v}_0, \boldsymbol{A}(T) \boldsymbol{v}_0,  \boldsymbol{A}(T)^2 \boldsymbol{v}_0, \ldots,  \boldsymbol{A}(T)^{n-1} \boldsymbol{v}_0 \},
\end{equation}
where $\boldsymbol{A}(\tau)^i v_0$ is obtained by direct calculation $\boldsymbol{v}_1 = \boldsymbol{A}(\tau)\boldsymbol{v}_0$, $\boldsymbol{v}_2 = \boldsymbol{A}(\tau)\boldsymbol{v}_1$, $\ldots$. Further eigenvalue calculations are carried out by standard numerical algebraic techniques, such as the Arnoldi method. The eigenvalues are obtained by \texttt{Nektar++} in the form:
\begin{equation}
\label{article30.eigenval_form}
\lambda(\tau) = m \cdot e^{\theta i},\\
\end{equation}
and if $m>1$ then the flow is unstable. The time-independent growth is $\sigma = \frac{ln(m)}{\tau}$ and the time-independent frequency is $\omega=\frac{\theta}{\tau}$. 

For eigenvalue problem it is possible to use a direct coupled approach developed by Sherwin\&Ainsworth\cite{sherwin2000unsteady} and also implemented in Nektar++. A weak form of the Stokes problem can be considered
\begin{equation}
\label{article30.stokes_problem}
\begin{aligned}
\left( \nabla \phi ,\frac{1}{Re}\nabla \boldsymbol{v}\right) &- \left( \nabla \cdot \phi, p\right) = \left(\phi, \boldsymbol{f} \right),\\
\left( q , \nabla \cdot \boldsymbol{v} \right) &= 0,
\end{aligned}
\end{equation}
where $\boldsymbol{v},\phi \in \boldsymbol{V}$, $p,q \in W$ and $\boldsymbol{v}$, $W$ are appropriate spaces for the velocity and pressure system to satisfy inf-sup condition. To complete system (\ref{article30.stokes_problem}) to the Navier-Stokes problem, the advection terms $\left(\boldsymbol{U}\cdot\nabla \right)\boldsymbol{v}+\left(\boldsymbol{v}\cdot\nabla \right)\boldsymbol{U}$ are included as force $\boldsymbol{f}$. Next, the Arnoldi method is applied.

\begin{figure}
\begin{center}
\includegraphics[width=0.4\textwidth]{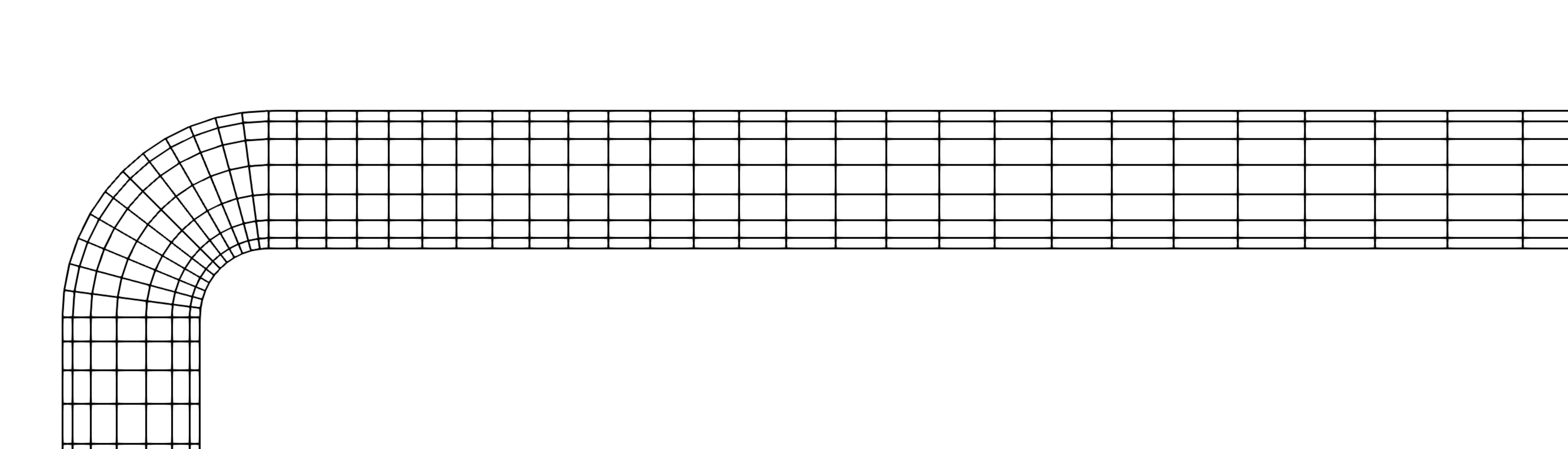}
\end{center}
\caption{Part of the mesh near the bend}\label{a30_Mesh}
\end{figure}

Figure \ref{a30_Mesh} shows the mesh that was used for the calculations. This mesh contains $685$ elements and is employed for flow at Reynolds number $Re=500\sim800$. For higher Reynolds numbers, meshes with more elements were used; for smaller $Re$, less detailed meshes are optimal.

\begin{figure*}
\begin{center}
\includegraphics[width=0.8\textwidth]{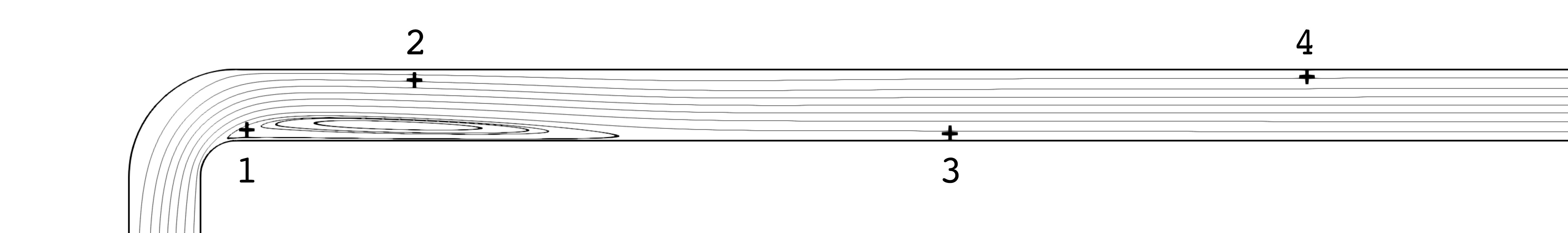}
\end{center}
\caption{The control points, $Re=600$}\label{a30_convPoints}
\end{figure*}

The steady-state flow was found by integrating equations (\ref{a30_NSEquation}) in time until a constant velocity was observed at the selected points and the time convergence was achieved in at least $9-10$ digits. Some of these points are marked with crosses in figure \ref{a30_convPoints}, the numbers increase from left (No 1) to right (No 4). Table \ref{a30_convRe600} shows the values of the horizontal velocity and the growth depending on the order of the approximation polynomials $p$. Velocity converges up to $6$ digits or $0.0001\%$ of the velocity scale $U$. Table \ref{a30_convRe600} also shows the $\sigma$ convergence for the leading (real) mode at $\beta=1.0$. The table also include $\sigma$ convergence for base flow time $T$. Most of the results below were obtained at $T=500-1000$ and the approximation order is $p_s=7$ for the base flow and $p_s=10-12$ for the eigenvalues. 

The eigenvalues and the base flow may depend on the length of the inlet and outlet branches of the channel. For verification, a series of calculations was performed with different $L_{in}$ and $L_{out}$ at $Re=710$, close to the critical Reynolds number. The results of these calculations are shown in the table \ref{a30_convRe710}. It was found that in the range $10 \leq L_{in} \leq 40$ and $40 \leq L_{out} \leq 120$ hold $\sigma_{max}-\sigma_{min}=2.5\cdot 10^{-6}$
This means that the effect the inlet and outlet lengths to the eigenvalues for $L_{in} \geq 10$ and $L_{out} \geq > 40$ is less than the calculations error. Such check was performed not only for $Re=710$, but also for larger Reynolds numbers.

\begin{table*}
\caption{The convergence of the growth $\sigma$ and the horizontal velocity $v_x$ at the control points, $Re=600$. Calculations were made on the mesh containing 685 elements exclude a column for which  the mesh size was 1341 element. $p_b$ was the order of approximation for the base flow, $p_s$ was the same for the stability calculations. Alone  $p$ means $p_b=p_s$. The tolerance was $10^{-8}$}
\label{a30_convRe600}
\scriptsize
\begin{tabular}{|c||c|c|c|c||c|c||c|c|c||}
\hline
p & Point 1       & Point 2 & Point 3 & Point 4 & $\sigma$ &  $\sigma$, $1341$ elem.&$T$& $p_b=7$, $p_s=7$&  $p_b=7$, $p_s=10$\\
\hline
5  &-0.0097407 &1.046623711 &0.27554782 &0.32007041 &-0.00362109 & -0.00341311 & 100 & -0.00349753 & -0.00350349\\
7  &-0.0085206 &1.046248285 &0.27560974 &0.32013706 &-0.00378192 & -0.00378138 & 250 & -0.00377620 & -0.00378212\\
10 &-0.0084460 &1.046219465 &0.27561238 &0.32013999 &-0.00378744 & -0.00378506 & 500 & -0.00378192 & -0.00378723\\
12 &-0.0084454 &1.046219463 &0.27561233 &0.32013996 &-0.00378747 & -0.00378506 & 1000& -0.00378192 & -0.00378724\\
\hline
\end{tabular}
\end{table*}

\begin{table*}
\caption{The influence of $L_{in}$ and $L_{out}$ on the growth $\sigma$ at $Re=710$}
\label{a30_convRe710}
\begin{tabular}{||c|c|c||c|c|c||c|c|c||}
\hline
$L_{in}$ & $L_{out}$ & $\sigma$ & $L_{in}$ & $L_{out}$ & $\sigma$ & $L_{in}$ & $L_{out}$ & $\sigma$\\
\hline
5  & 60 & -0.00005779 & 10  & 40  & 0.00012109 & 10 & 60  & 0.00012108\\
10 & 60 &  0.00012108 & 10  & 60  & 0.00012108 & 20 & 80  & 0.00012352\\
20 & 60 &  0.00012351 & 10  & 80  & 0.00012111 & 30 & 100 & 0.00012353\\
30 & 60 &  0.00012351 & 10  & 100 & 0.00012109 & 40 & 120 & 0.00012116\\
\hline
\end{tabular}
\end{table*}

{\AA}kervik et al. \cite{aakervik2006steady} describe a selective frequency damping method that can converge to an unstable flow. This is a replacement of the Newton method, and it is easier to use since the method does not require a high quality initial condition. However, the selective frequency damping method requires a very large computational cost.  

Jordy et al. in their article \cite{jordi2015adaptive} describe an adaptive selective frequency dumping method, which applies one-dimensional reduced model to find these parameters. The adaptive selective frequency damping method is implemented in the Nektar++ framework \cite{jordi2014encapsulated, jordi2015adaptive}. The complete numerical calculations set up and the convergence analysis are presented in article\cite{proskurin2020mathematical}.

A nonlinear form of the advection terms
\begin{equation}
\label{a30_nonLin}
\left(\boldsymbol{V}\cdot\nabla \right) \boldsymbol{V} =  \left(\boldsymbol{U}\cdot\nabla \right) \boldsymbol{v}+\left(\boldsymbol{v}\cdot\nabla \right) \boldsymbol{U}+\left(\boldsymbol{v}\cdot\nabla \right) \boldsymbol{v}
\end{equation}
can be used to split the base flow and the disturbance dynamics calculations. Appropriate modification of the Nektar++ code was reported in  \cite{proskurin2019numerical,proskurin2019evolution}. This approach allows the study of the nonlinear disturbances near the stabilized base flow.

\section{Flow regimes}\label{a30_flowRegimes}

\begin{figure*}[ht]
\begin{center}
\tiny
% \begin{tabu}{X[0.5,rm]X[12,lm]}
\begin{tabular}{cc}
(a)
&
\includegraphics[width=0.9\textwidth, viewport= 5cm 0cm 60cm 8cm,clip]{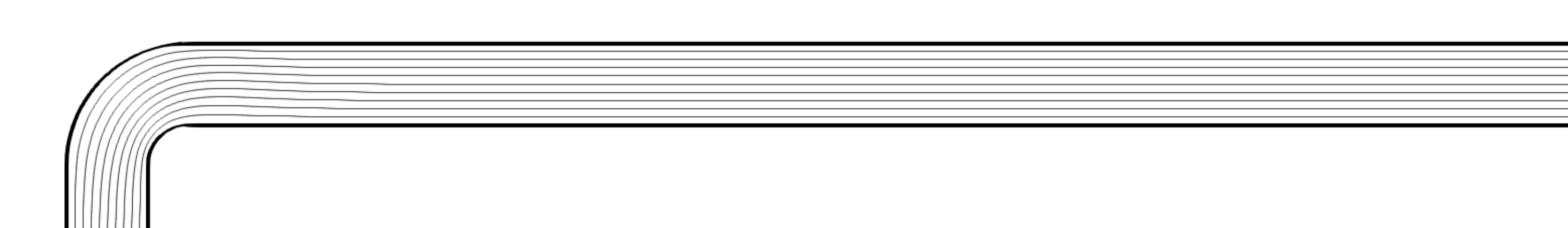}
\\
(b)
&
\includegraphics[width=0.9\textwidth, viewport= 5cm 0cm 60cm 8cm,clip]{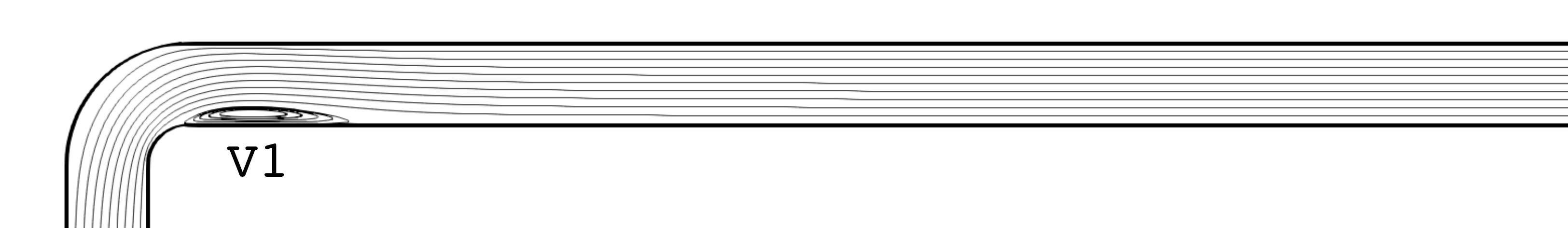}
\\
(c)
&
\includegraphics[width=0.9\textwidth, viewport= 5cm 0cm 60cm 8cm,clip]{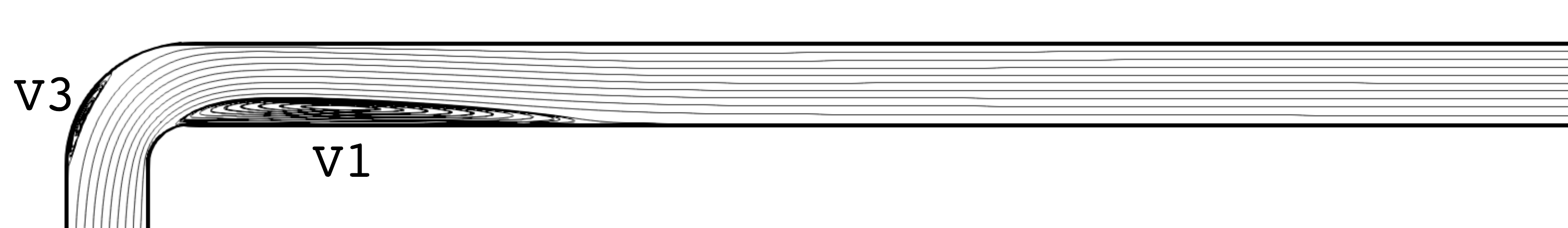}
\\
(d)
&
\includegraphics[width=0.9\textwidth, viewport= 5cm 0cm 60cm 9cm,clip]{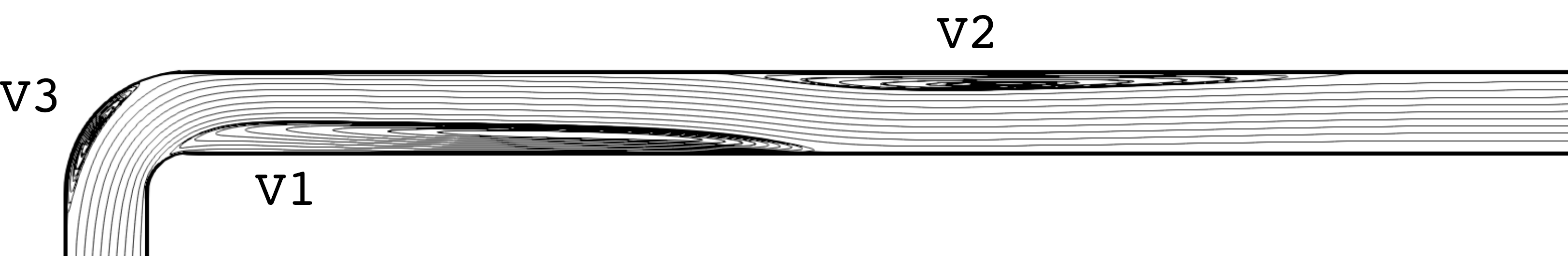}
\\
(e)
&
\includegraphics[width=0.9\textwidth, viewport= 5cm 0cm 60cm 8cm,clip]{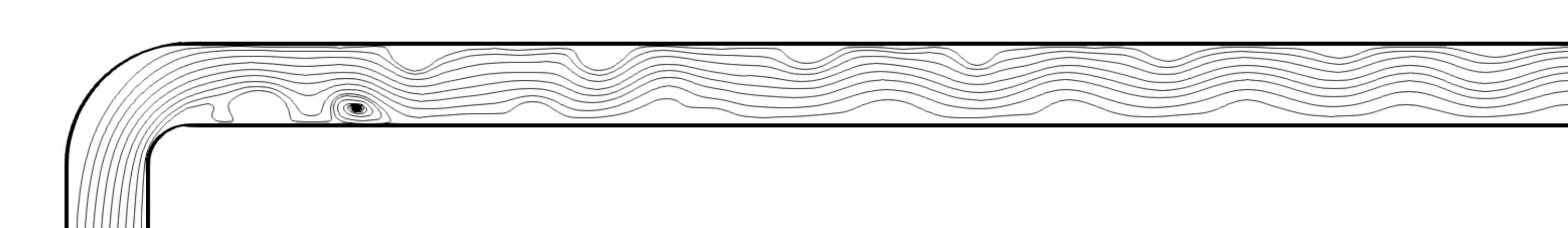}
\\
\end{tabular}
\end{center}
\caption{The base flow patterns: $Re=20$(a),$Re=200$(b),$Re=500$(c),$Re=1300$(d)(the steady-state flow),$Re=1300$(e)(the pulsating flow)}\label{a30_Patterns2D}
\end{figure*}

Figure \ref{a30_Patterns2D} shows streamlines at $\delta=1$ and $Re=20$, $200$, $500$, $1300$. The flow is laminar at small Reynolds numbers and its streamlines are parallel (see \ref{a30_Patterns2D}(a)). When the Reynolds number increases, vortices $V1$, $V3$ and $V2$ (see \ref{a30_Patterns2D}(b,c,d)) appear. For relatively high Reynolds numbers it is possible for two different flow patterns to exist: the steady-state and the pulsating. The pulsating pattern is shown in figure \ref{a30_Patterns2D}(e). One of the two patterns can exist depending on the conditions that are described later in the article.

\begin{figure}
\begin{center}
\includegraphics[width=0.5\textwidth, viewport= 1cm 0cm 8cm 2cm,clip]{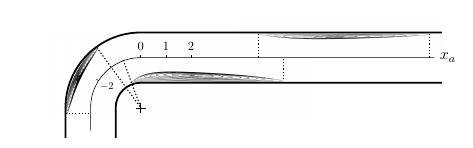}\\
\includegraphics[width=0.5\textwidth]{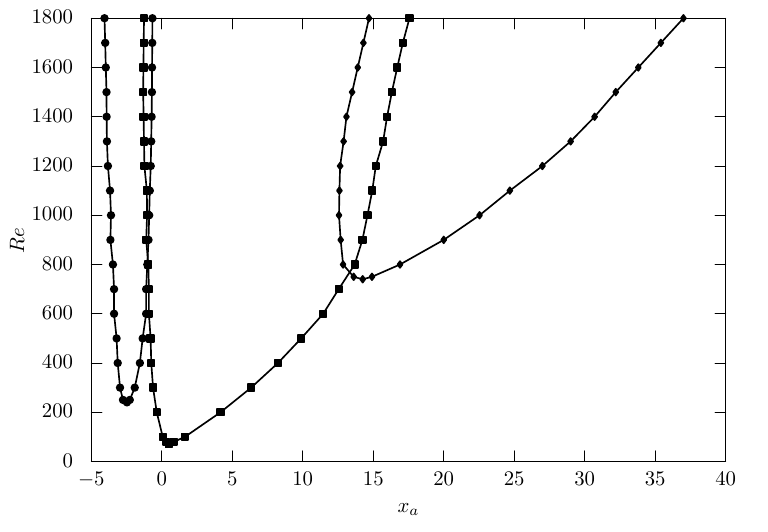}
\end{center}
\caption{Flow diagram, $\delta=1$, the circles indicate the appearance of $V3$ recirculation bubble, rectangles -- $V1$, diamonds -- $V2$, the dashed line is rough lower border for the pulsating flow origin. The top panel shows a curvilinear coordinates with an axle $x_a$ and projections of the separation points on the axle}\label{a30_FlowDiagramDelta1}
\end{figure}

Figure \ref{a30_FlowDiagramDelta1} presents a vortex diagram in the bend. $V3$ vortex area is marked by circles, $V1$ by squares, and $V2$ by diamonds. To calculate the end position of the vortices, a curved coordinate system is introduced, the axis  where $x_a$ is aligned along the axis of the channel. The ends of the vortices are projected onto it, as shown on top panel in figure \ref{a30_FlowDiagramDelta1} using dotted lines. 

Matsumoto et al. \cite{matsumoto2016two} found the vortices $V1$, $V2$ in the sharp bend. Zhang\&Poth\'erat and Sapardi et al. \cite{zhang2013influence,sapardi2017linear}  found vortices $V1$, $V2$, and $V3$ in the 180-degree sharp bend, where $V1$ vortex has a great influence on the flow's characteristics. Articles \cite{matsumoto2016two, zhang2013influence,sapardi2017linear} also describe the appearance of the pulsating flow.

\section{The pulsatile flow}\label{a30_2DStabNonlin}

This section presents the results of a study of the pulsations that are observed in the outlet branch at moderate  Reynolds numbers. Matsumoto et al. \cite{matsumoto2016two} have found such non-stationary patterns in the sharp bend. Zang\&Poth{\'e}rat and Sapardi et al. \cite{zhang2013influence, sapardi2017linear} found such vortex shedding in the sharp 180-degree bend.  Sapardi et al. have found the large hysteresis for the critical Reynolds number that depends from the initial conditions. They found for a special case, that $Re'_{\ast}\approx 1150$ for calculations from rest, or the lower Reynolds number steady-state velocity field, and $Re'_{\ast}=743$ for calculations from the unsteady initial condition. The apostrophe here means that they calculated the critical Reynolds number by the channel width and $Re = 0.5 Re'$. Yamasita et al.\cite{yamashita1986fluid} experimentally observed the pulsatile flow in a sharp two-dimensional bend at $Re>172.5-750$ (Reynolds number converted to this article parameters).

During routine calculations for this investigation, it had observed that an increase in the quality of the approximation, i.e., a decrease in the time step, an increase in the order of approximation $p$, or the number of mesh elements, suppresses the vortex shedding. Therefore, it is possible to assume that these pulsations arise due to the instability of the flow under the influence of perturbations imposed by the numerical scheme noise.

\begin{figure*}
\begin{center}
\begin{tabular}{cc}
\includegraphics[width=0.4\textwidth]{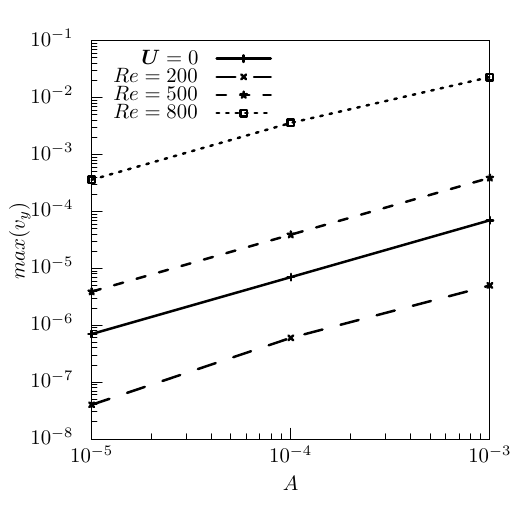}
&
\includegraphics[width=0.4\textwidth]{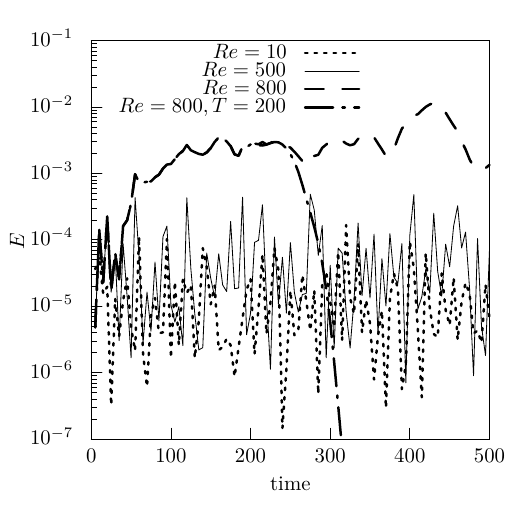}
\\
(a) & (b)\\
\end{tabular}
\end{center}
\caption{The response for the random noise as the vertical velocity $v_y$ maximal amplitude in the outlet branch centre line for the motionless fluid and $Re=200$, $500$, $800$(a); The disturbance energy for $Re=10$, $500$, $800$ for the permanent noise amplitude $A=10^{-3}$ and, for the case $Re=800$, the noise had been switched off at $t=200$(b)}
\label{a30_RandForceResponce}
\end{figure*}

To prove this hypothesis, the perturbation was imposed as the force $\boldsymbol{F}$ (see (\ref{a30_NSEquation})) of the random amplitude $A$ for the two-dimensional base flow. The values of this force were updated at each time step. Figure \ref{a30_RandForceResponce}(a) shows the flow response as the dependencies of the maximum vertical velocity amplitude on the centre line of the outlet branch from the noise amplitude. The curve $\boldsymbol{U} = 0 $ shows the response in the case of the rest fluid, when there is no gain due to the instability. In the case of $Re = 200$ this response is smaller than for the rest fluid and the perturbations do not grow because disturbance flow out from the channel. In the cases $Re=500$ and $Re=800$ the perturbations are amplified, and when $Re=800$, the motion amplitude is  $10^2$ times greater compared to the rest fluid. 

Figure \ref{a30_RandForceResponce}(b) shows graphs of the kinetic energy of the perturbation under the noise with the amplitude $A=10^{-3}$. At $Re=10$, the perturbation energy is slightly lower than at $Re=500$, and for these cases the energy of the forced motion remains relatively small. At $ Re=800$, the energy of the perturbed motion at the initial step has the same level as at $Re=500$. At $t\approx 50$ this energy increases at least $10$ times and the oscillatory motion continues at this energy level. If the noise was been switched off at time $t=200$, the pulsating motion faded. Thus, the pulsations must be supported by permanent external impulsion.

\begin{figure*}
\begin{center}
\includegraphics[width=1.0\textwidth]{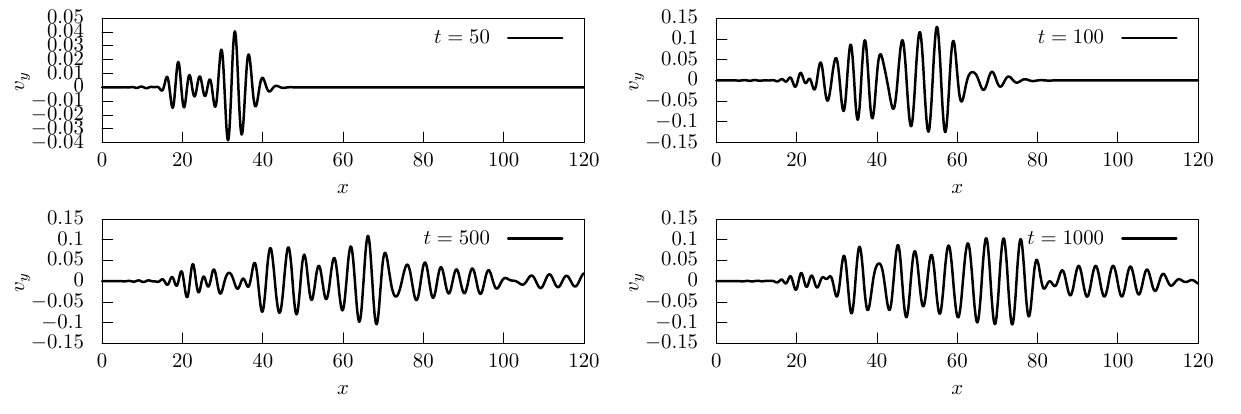}
\end{center}
\caption{The graphs of the vertical velocity $v_y$ at the outlet branch centreline for $t=50$, $100$, $500$, $1000$, $Re=1000$}\label{a30_VyOnL}
\end{figure*}

Figure \ref{a30_VyOnL} presents the vertical velocity on the outlet branch centre line at several consecutive moments in time after the noise is turned on. Each period of the graph corresponds to one vortex. Vortices are arising near the trailing edge of $V2$. The velocity amplitude of each vortex increases downstream and reaches a maximum at $x>50$. Then each vortex moves with its own amplitude until it leaves the channel. The amplitude of the vortices along the longitudinal coordinate $x$ fluctuates in some way, not quite exactly periodically. This means that the vortex path is modulated in amplitude in the longitudinal direction, in other words, a sequence of vortex packets forms in the channel. 

It is known that some flows can amplify disturbances. Boujo\&Gallaire\cite{boujo2015sensitivity} studied this phenomenon in the flow over the backward-facing step  and found that the steady-state flow is stable, but disturbances can increase in space and time  as a result of non-normal effects. As is clear from the above results, this amplification are observed in the bent flow. In context of this investigation, this phenomenon is completely eliminated by increasing the accuracy of calculations or using the selective frequency dumping method.

\section{Stability of the two-dimensional flow}\label{a30_2DStab}

This section considers the stability of the two-dimensional perturbations at $\delta=1$. The base steady-state flow was calculated up to $Re\sim 1200$ by time-integrating the equations (\ref{a30_NSEquation}). For larger Reynolds numbers, the adaptive selective frequency dumping method described above was applied. With this method, it was possible to calculate the steady-state flow up to $Re=1900$. The direct solver was used for eigenvalues together with the SFD baseflow calculations, because in this case the time-dependent scheme (\ref{article30.stab_eigproblem}) is unstable for the same reason as for the baseflow.

\begin{figure*}
\begin{minipage}{0.49\textwidth}
\includegraphics[width=0.8\textwidth]{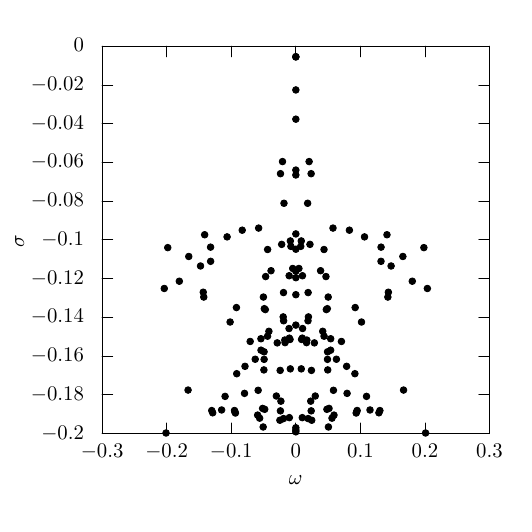}
\caption{The eigenvalues at $Re=1800$, $\delta=1$}\label{a30_spectra2DRe1800Delta1}
\end{minipage}
\begin{minipage}{0.49\textwidth}
\includegraphics[width=0.8\textwidth]{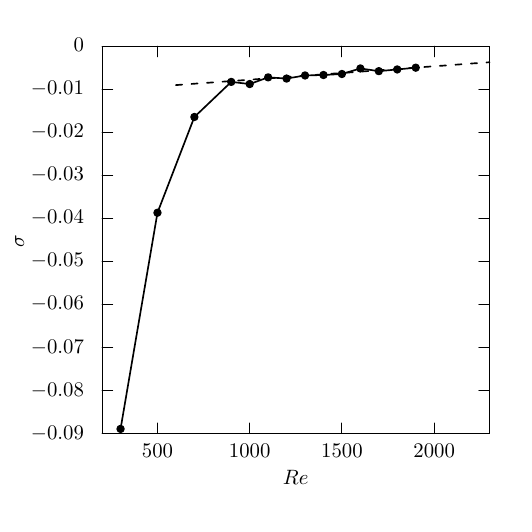}
\caption{The growth $\sigma$ form $Re$ for the 2D disturbances, the dashed line is the least squares approximation for $Re > 900$}\label{a30_growthFromRe2D}
\end{minipage}
\end{figure*}

\begin{figure*}[bt]
\begin{center}
% \begin{tabu}{X[0.5,rm]X[12,lm]}
\begin{tabular}{cc}
(a)&
\includegraphics[width=0.9\textwidth]{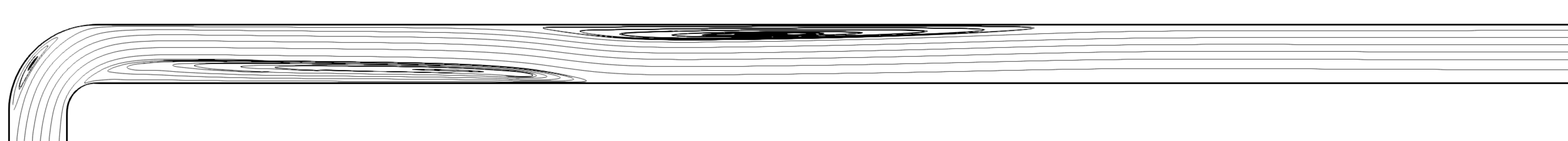}
\\
(b)
&
\includegraphics[width=0.9\textwidth]{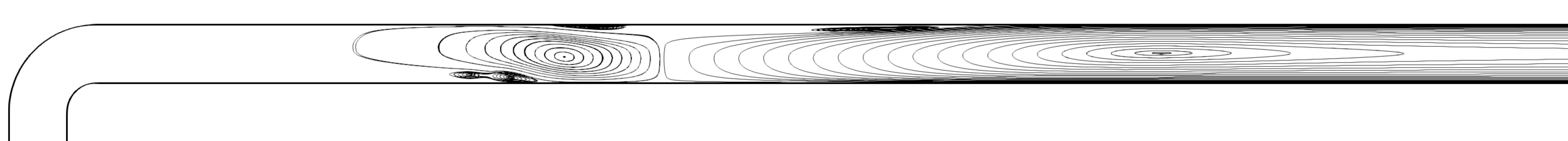}
\\
(c)
&
\includegraphics[width=0.9\textwidth]{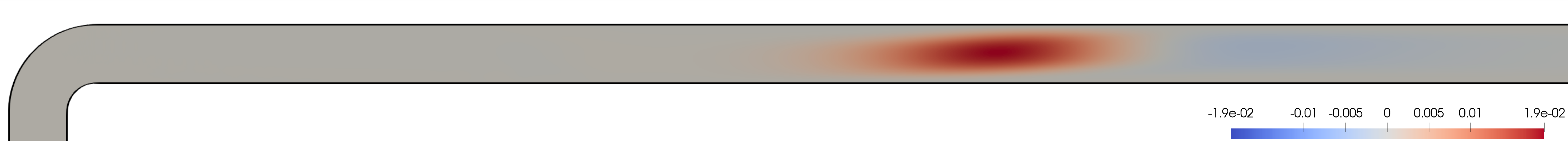}
\\
\end{tabular}
\end{center}
\caption{The stabilized flow streamlines at $Re=1600$ (a), its leading eigenfunction (b) and the vertical velocity amplitude (c)}\label{a30_2Dstab1600}
\end{figure*}

In Figure \ref{a30_spectra2DRe1800Delta1} presents the eigenvalues for $Re=1800$ and $\delta=1.0$.
The most dangerous is the monotonic mode. Figure \ref{a30_growthFromRe2D} shows the dependencies of $\sigma$ from $Re$. In this graph, all $\sigma$ values are lower than zero, so the two-dimensional perturbations are stable up to $Re=1900$. A dashed line represents an approximation $\sigma(Re)$ for $900 \leq Re \leq 1900$ using the least squares method. The appropriate critical Reynolds number is $Re_{\ast} = 3495$. Figure \ref{a30_2Dstab1600} shows the streamlines of the base flow(a) and the leading mode(b) at $Re=1600$, and the vertical velocity amplitude(c). Thus, the most dangerous mode is localized at downstream the recirculation bubble $V2$.

% \begin{figure}
% \begin{center}
% \includegraphics[width=0.45\textwidth]{a30_spectra2DRe1800Delta1.pdf}
% \end{center}
% \caption{The eigenvalues at $Re=1800$, $\delta=1$}\label{a30_spectra2DRe1800Delta1}
% \end{figure}
% 
% \begin{figure}
% \begin{center}
% \includegraphics[width=0.5\textwidth]{a30_growthFromRe2D.pdf}
% \end{center}
% \caption{The growth $\sigma$ form $Re$ for the 2D disturbances, the dashed line is the least squares approximation for $Re > 900$}\label{a30_growthFromRe2D}
% \end{figure}

\section{Linear stability}\label{a30_3DStab}

\begin{figure*}
\begin{center}
\begin{tabular}{cc}
\includegraphics[width=0.45\textwidth]{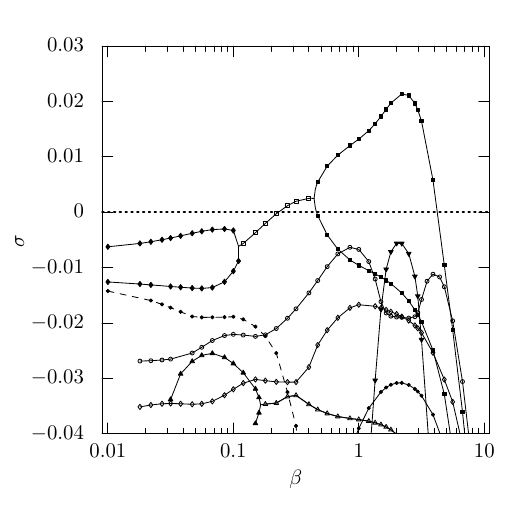}&
\includegraphics[width=0.45\textwidth]{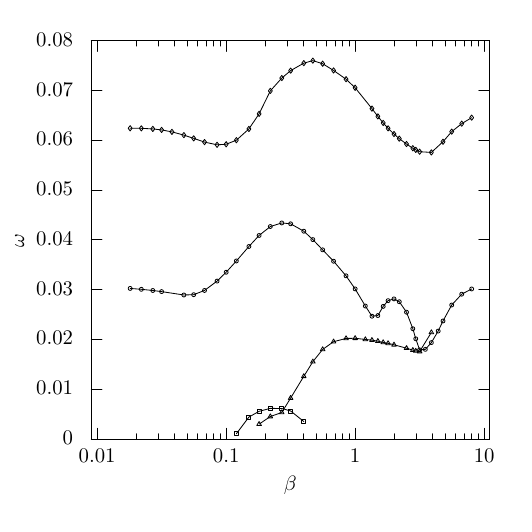}\\
(a)&(b)
\end{tabular}
\end{center}
\caption{The eigenvalues at $Re=1400$ as function of spanwise wavenumber $\beta$: growth $\sigma$ (a) and frequency $\omega$ (b). Only positive frequencies are shown here.}\label{a30_spectraDelta1modes}
\end{figure*}

This section summarizes the results of studying three-dimensional linear stability in the bent channel. Figure \ref{a30_spectraDelta1modes} presents the growth $\sigma$(a) and frequency $\omega$(b) at $Re=1400$ as function of the spanwise wavenumber $ \beta=\frac{2\pi}{L}$. Solid markers signify real modes, and hollow markers -- complex modes. The graph \ref{a30_spectraDelta1modes} is plotted in the range $0.01 < \beta < 10$, outside of which the flow is stable. Small values of $\beta$ correspond to perturbations with a very large spanwise wavelength, and, in limit case  $\beta \rightarrow0$, fit to the two-dimensional case. In the case of large $\beta$ spanwise wavelength of the perturbation is short. The leading real mode is indicated by solid rectangles. It has a maximum at $ \beta \approx 2.2$. In the range of $0.11<\beta<0.44$, the leading mode is complex. For sufficiently large Reynolds numbers, this mode can be unstable. The spatial structure of unstable modes will be considered below. Solid diamonds denote a real mode that is the leading at $\beta<0.11$, and this mode is stable. Empty circles and diamonds mark complex modes. Their amplitude sufficiently large in the downstream area. Also, the graph shows several minor stable modes.

\begin{figure*}
\begin{minipage}{0.45\textwidth}
\includegraphics[width=0.9\textwidth]{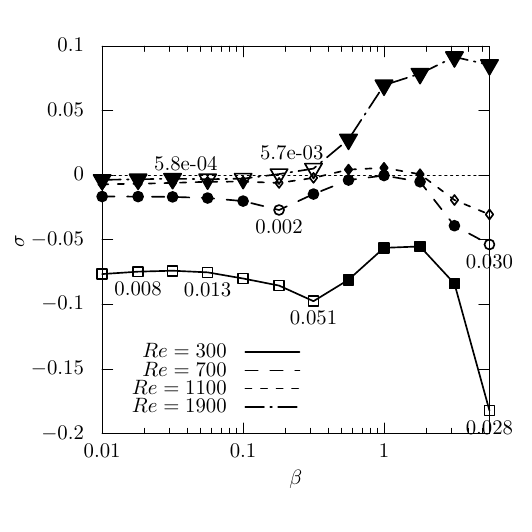}
\caption{Leading growths $\sigma$ at $Re=300$, $700$, $1100$, $1900$ as function of the spanwise wavenumber $\beta$. Solid points represent real eigenvalues, hollow points represent complex eigenvalues. Numbers near selected points show value of the frequency $\omega$. Lines connect leading eigenvalues for each Reynolds number.}\label{a30_spectraReFromLDelta1}
\end{minipage}
\hspace{3mm}
\begin{minipage}{0.45\textwidth}
\includegraphics[width=0.9\textwidth]{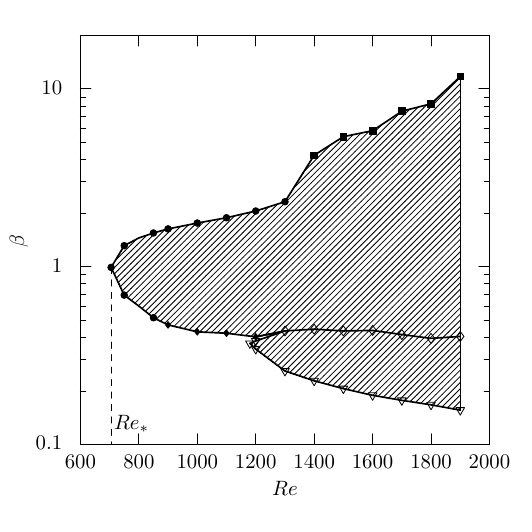}
\caption{Neutral stability curve for the bent flow. Filled region includes parameters' values for which the flow is unstable.}\label{a30_neutralDelta1}
\end{minipage}
\end{figure*}

Figure \ref{a30_spectraReFromLDelta1} shows the dependencies of the growth $\sigma$ on the spanwise wavenumber $\beta$ at $Re=300$, $700$, $1100$, $1900$. Each curve is made up of the values of $\sigma$ that correspond to the major mode. The real eigenvalues in this figure are marked by solid symbols, and the complex ones by hollow symbols. At $Re=300$, the curve has local maxima at $ \beta \approx 0.03$ and $ \beta \approx 1.0$, the first of which corresponds to the complex mode, and the second to the real mode. When $\beta$ decreases, the frequency also decreases. As the Reynolds number rises, the real mode first becomes unstable and there is a finite range of the spanwise wavenumbers for unstable modes. With a further increase in the Reynolds number, the left boundary of this interval is determined by periodical perturbations. Periodic modes also dominate at large $\beta$, but they are stable.

A neutral curve is presented in Figure \ref{a30_neutralDelta1}. This figure shows a area of instability for $Re < 1900$. The border points on the curve were obtained by detecting the zero-growth point by varying $\beta$ at a fixed Reynolds number. The unstable modes exist in the shaded area. The smallest value of the Reynolds number $Re$ at which instability is observed is the critical Reynolds number $Re_{\ast}$, which is associated with the critical wave number $\beta_{\ast}$. Empty diamonds indicate the boundary where the leading modes change from the real mode to the complex mode. Empty triangles denote the neutral boundary for the complex mode.

% \begin{figure}
% \begin{center}
% \includegraphics[width=0.45\textwidth]{a30_spectraReFromLDelta1.pdf}
% \end{center}
% \caption{Leading growths $\sigma$ at $Re=300$, $700$, $1100$, $1900$ as function of the spanwise wavenumber $\beta$. Solid points represent real eigenvalues, hollow points represent complex eigenvalues. Numbers near selected points show value of the frequency $\omega$. Lines connect leading eigenvalues for each Reynolds number.}\label{a30_spectraReFromLDelta1}
% \end{figure}
% 
% \begin{figure}
% \begin{center}
% \includegraphics[width=0.45\textwidth]{a30_Neutral.pdf}
% \end{center}
% \caption{Neutral stability curve for the bent flow. Filled region includes parameters' values for which the flow is unstable.}\label{a30_neutralDelta1}
% \end{figure}

\begin{figure*}[h!]
\begin{center}
\begin{tabular}{cc}
(a)&(aa)\\
\includegraphics[width=0.25\textwidth,viewport= 15cm 0cm 60cm 19cm,clip]{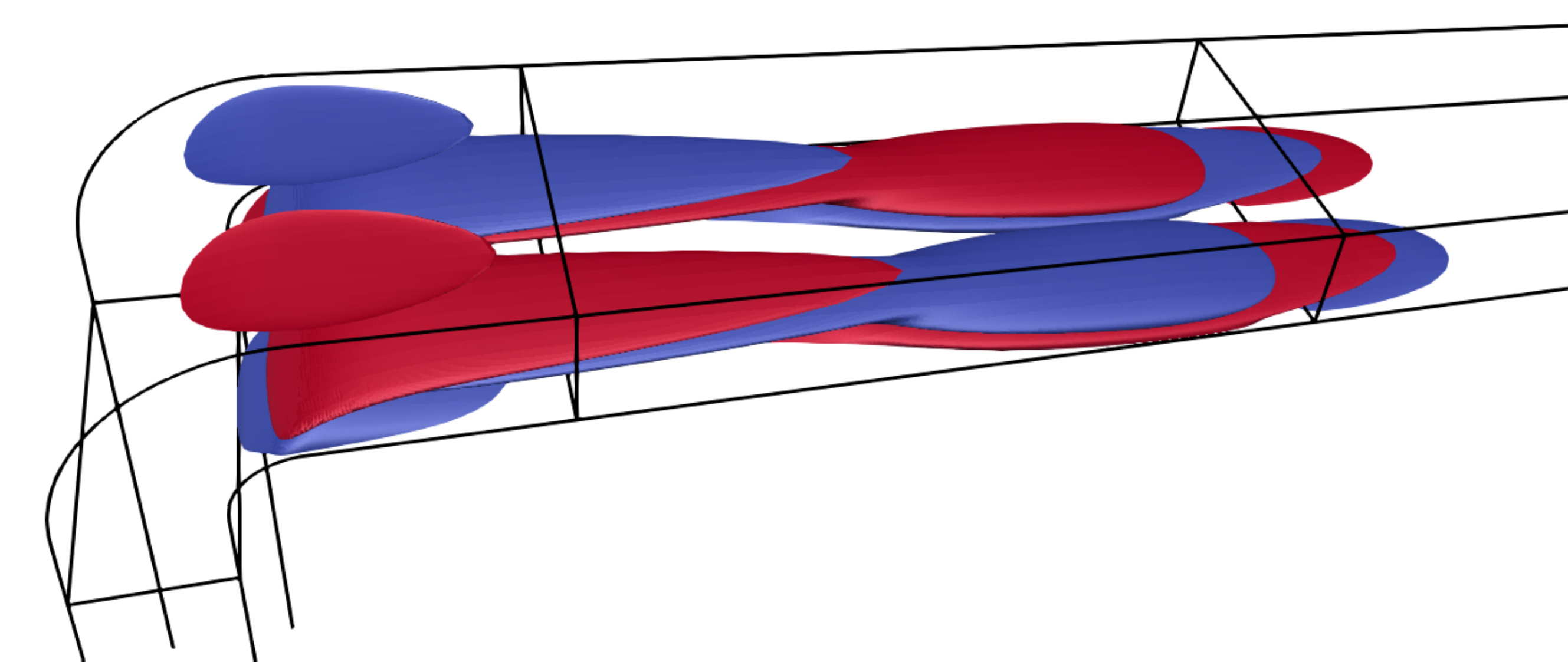}
&
\includegraphics[width=0.7\textwidth]{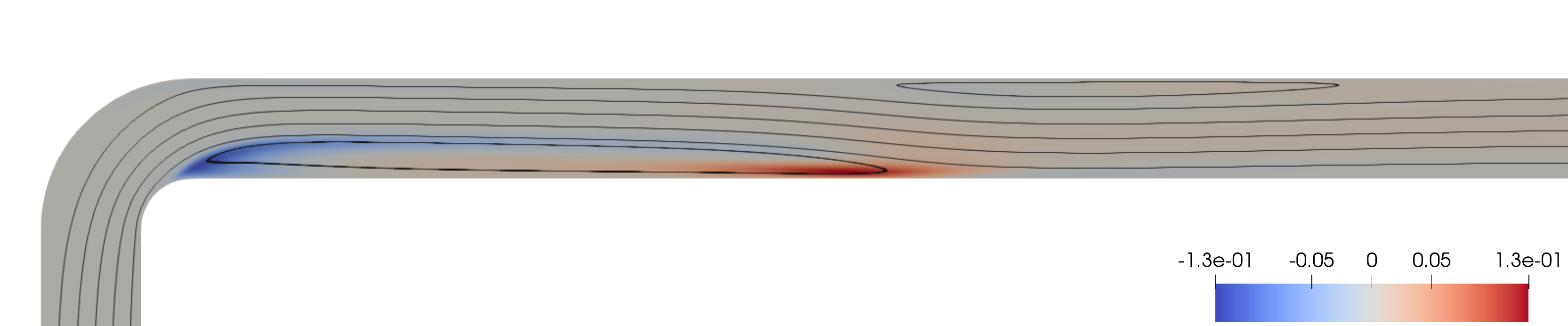}
\\
\hline
\multicolumn{2}{l}{(b) \hspace{1cm} \parbox[c]{0.5\textwidth}{\includegraphics[width=0.5\textwidth,viewport= 27cm 5cm 75cm 25cm,clip]{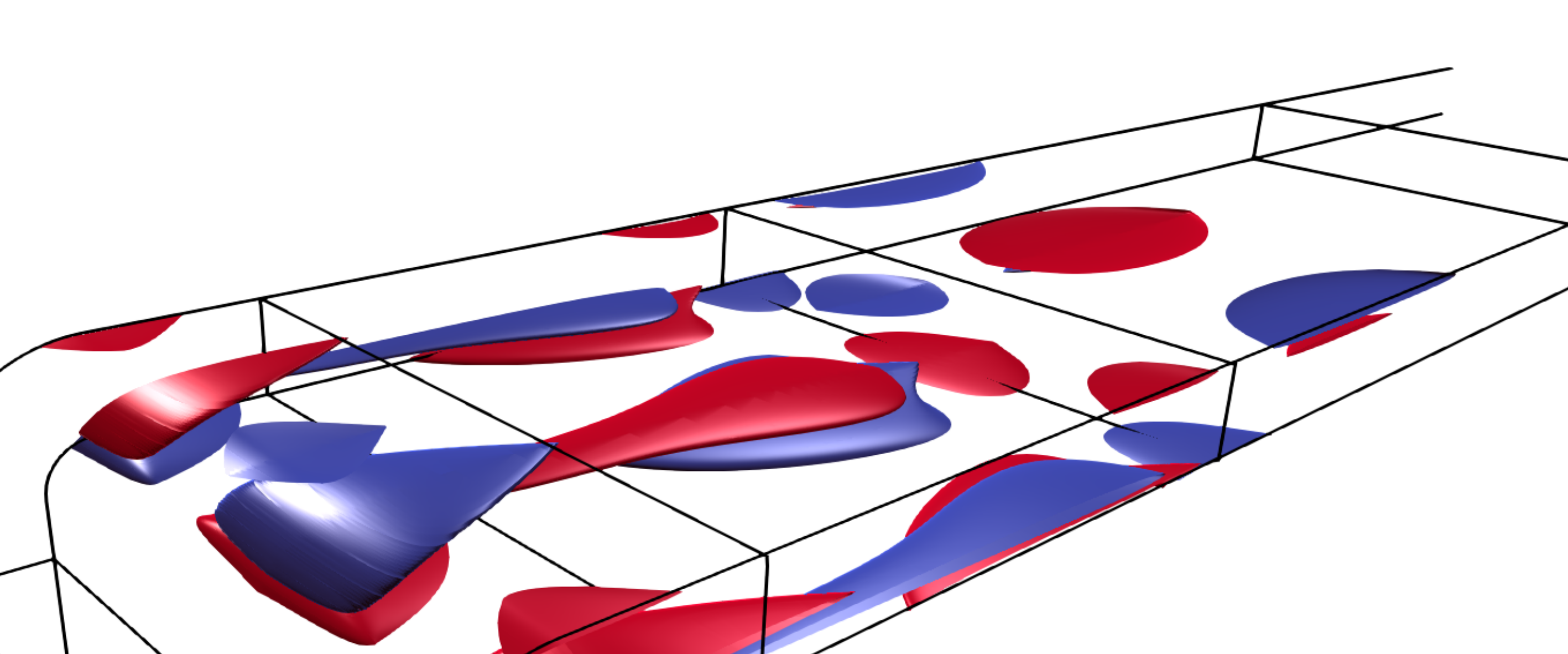}}}\\
\vspace{5mm}&\\
\multicolumn{2}{l}{(bb) \hspace{0.1cm} \parbox[c]{0.93\textwidth}{\includegraphics[width=0.93\textwidth]{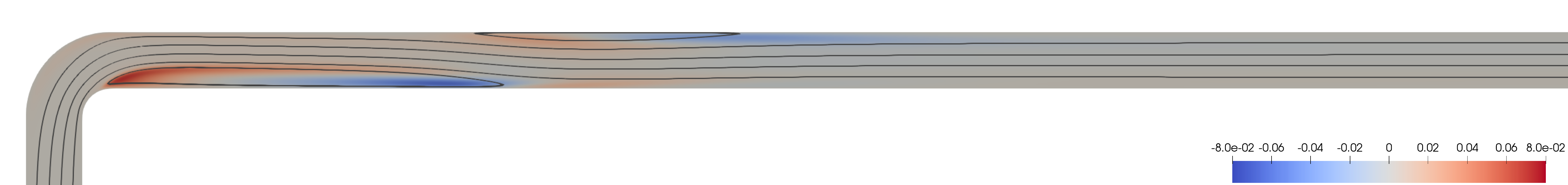}}}\\
\hline
\vspace{5mm}&\\
(c)&(cc)\\
\includegraphics[width=0.25\textwidth,viewport= 26cm 5cm 70cm 25cm,clip]{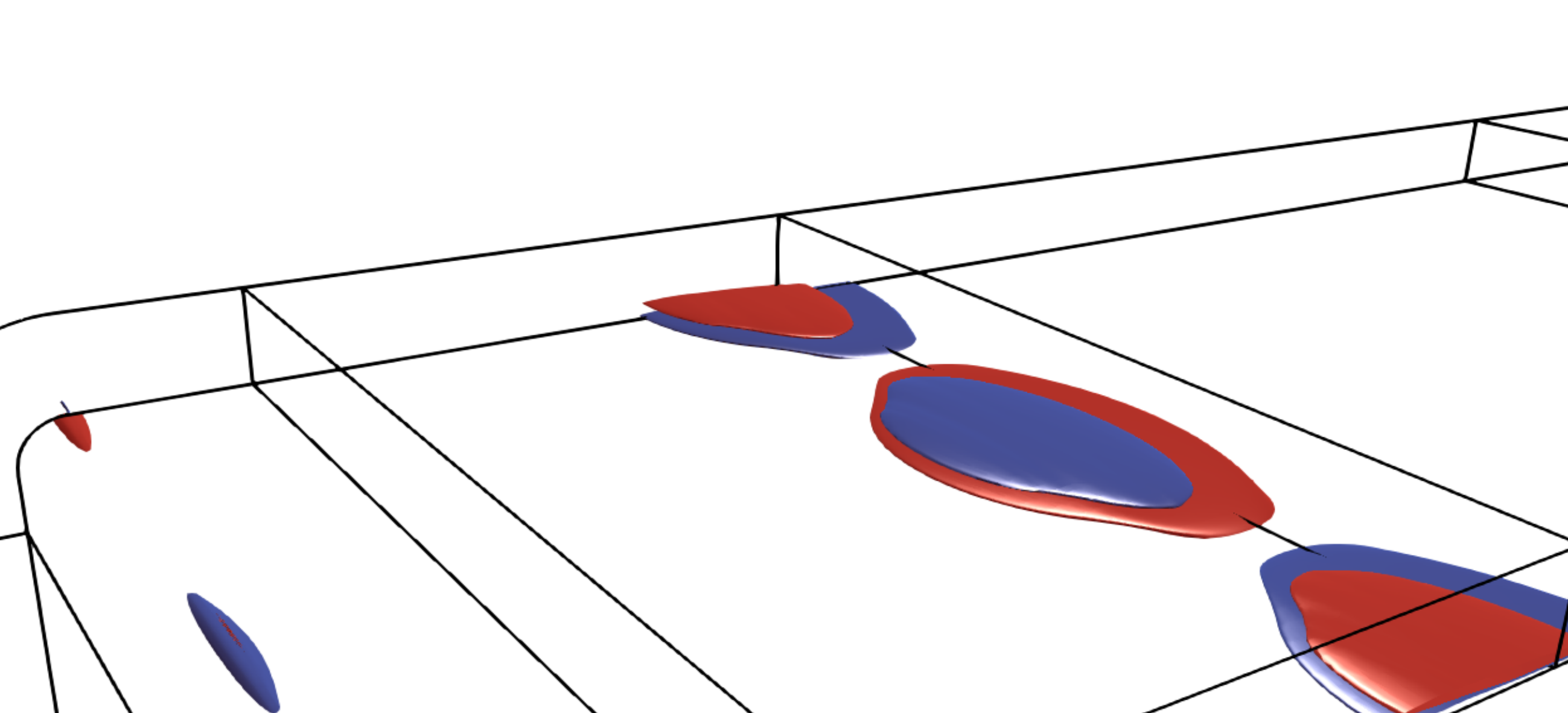}
&
\includegraphics[width=0.7\textwidth]{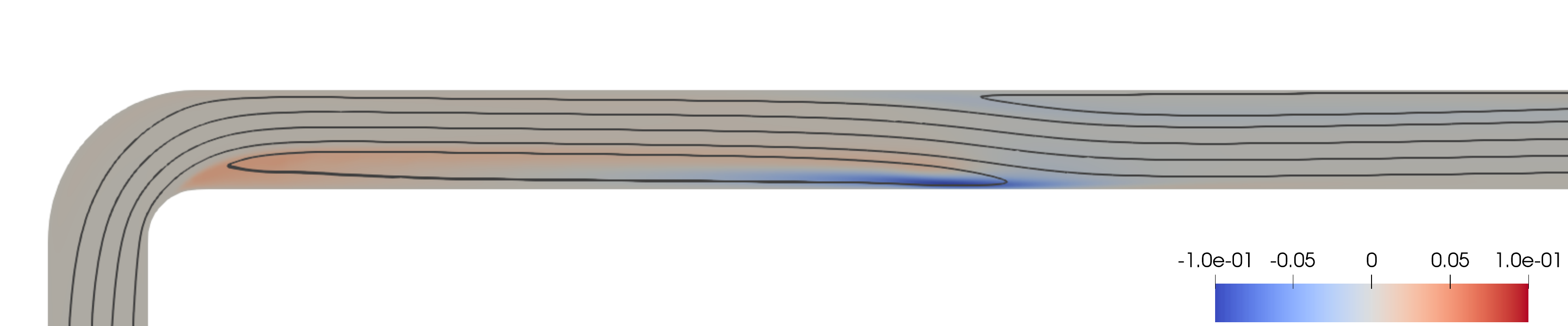}
\\
\hline
\vspace{5mm}&\\
\multicolumn{2}{l}{(d) \hspace{1cm} \parbox[c]{0.5\textwidth}{\includegraphics[width=0.5\textwidth,viewport= 26cm 5cm 100cm 25cm,clip]{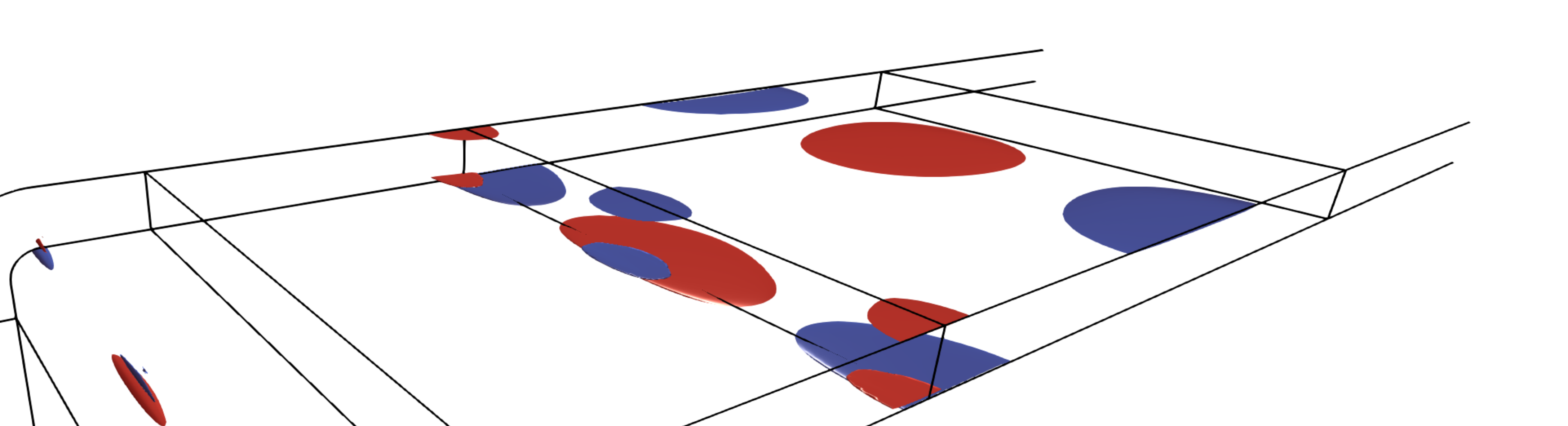}}}\\
\multicolumn{2}{l}{(dd) \hspace{0.1cm} \parbox[c]{0.93\textwidth}{\includegraphics[width=0.93\textwidth]{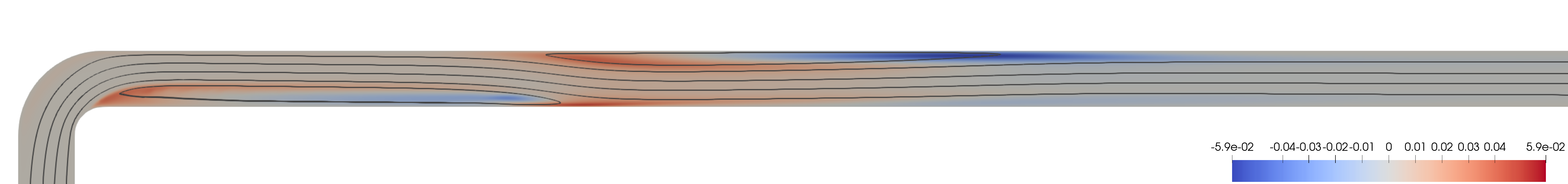}}}\\
\vspace{5mm}&\\
\hline
(e)& (ee)\\
\parbox[c]{0.25\textwidth}{\includegraphics[width=0.25\textwidth,viewport= 25cm 5cm 60cm 25cm,clip]{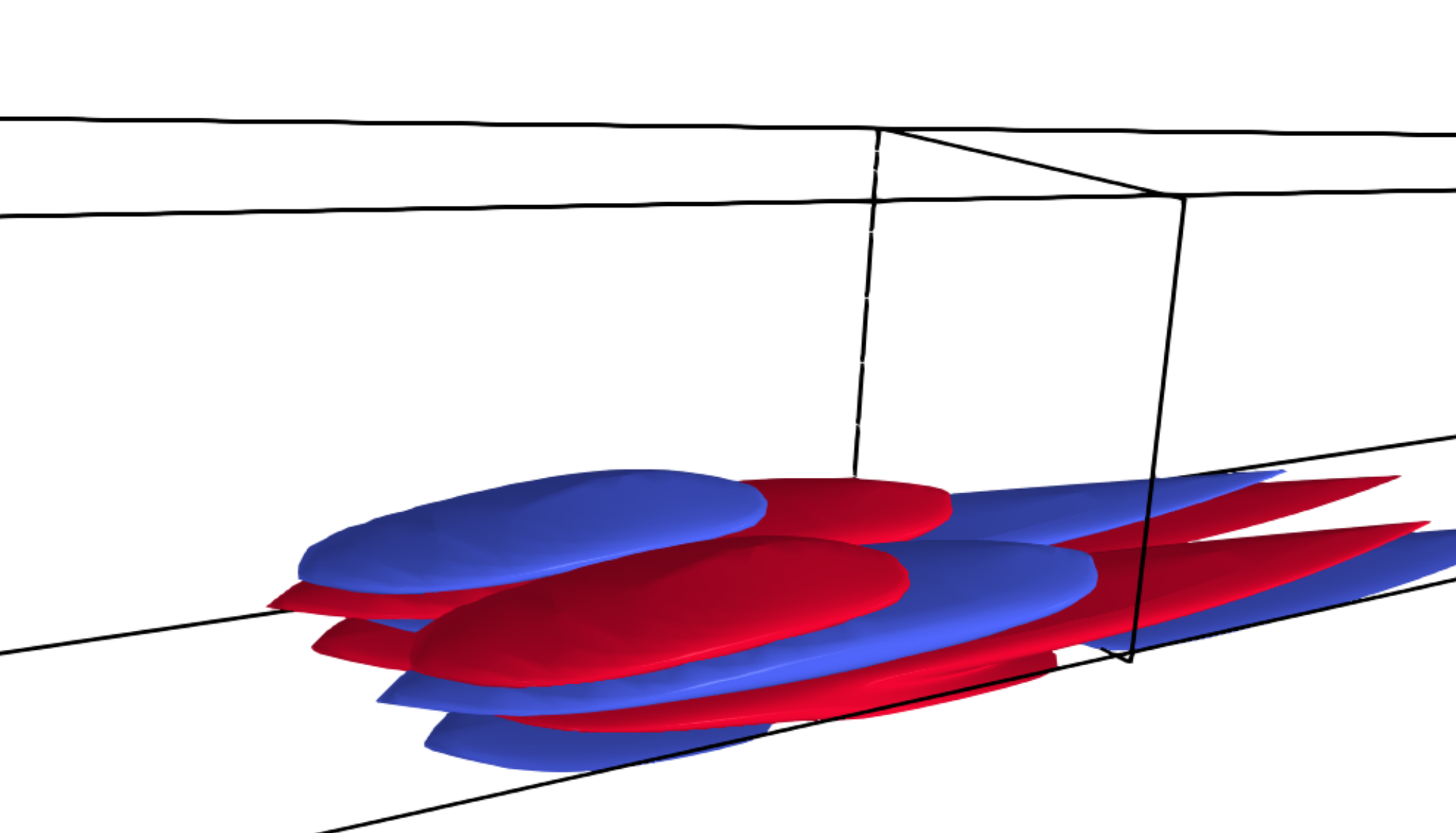}}
&
\parbox[c]{0.7\textwidth}{\includegraphics[width=0.7\textwidth]{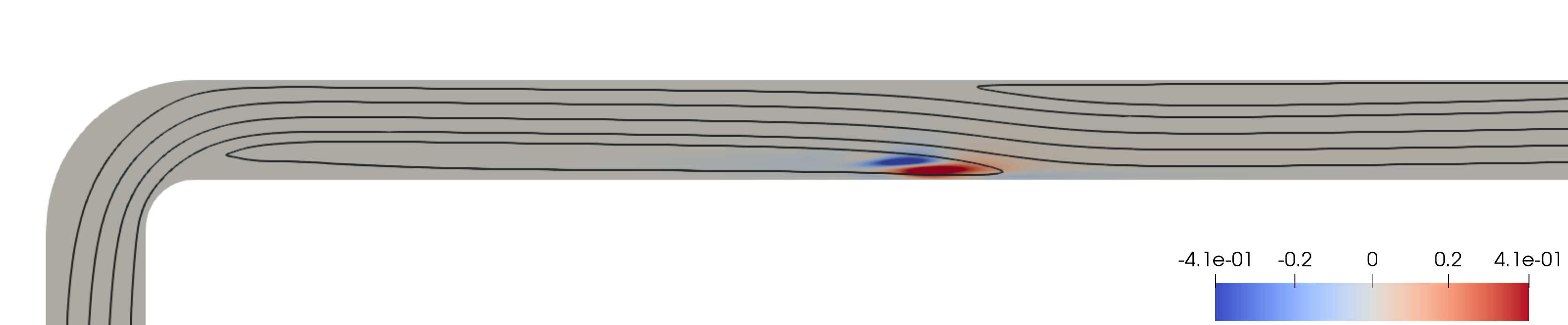}}
\\
\end{tabular}
\end{center}
\caption{The unstable modes visualisation: the bubble mode vorticity isosurfaces $\pm0.1$ (a, positive value is red, negative is blue) and $v_z$ amplitude  over the base flow streamlines, section at the maximal amplitude (aa) at $Re=1100$, $\beta=1$; the coupled mode vorticity isosurfaces $\pm0.15$ (b) and $v_z$ amplitude  over the base flow streamlines (bb) at $Re=1000$, $k=0.45$; the coupled complex mode vorticity isosurfaces $\pm0.3$ (real part of eigenfunction -- c, complex part -- d) and $v_z$ amplitude  over the base flow streamlines (real part -- cc, complex part -- dd) at $Re=1600$, $\beta=0.3$; the bubble tail mode vorticity isosurfaces $\pm0.4$ (a) and $v_z$ amplitude  over the base flow streamlines, section at the maximal amplitude (aa) at $Re=1600$, $\beta=3$}\label{a30_eigenfun3D}
\end{figure*}

\begin{figure*}
\begin{center}
\begin{tabular}{ccc}
\includegraphics[width=0.33\textwidth]{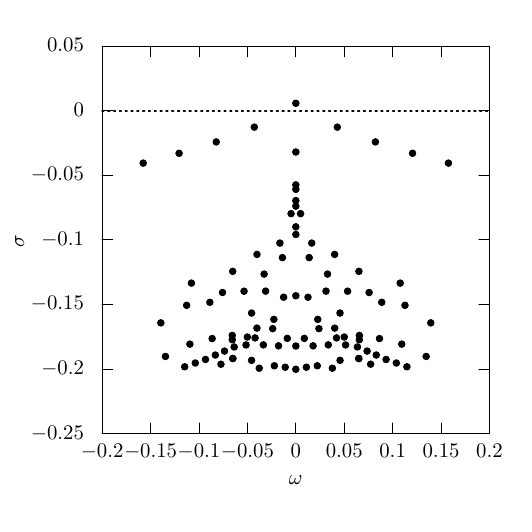}&
\includegraphics[width=0.33\textwidth]{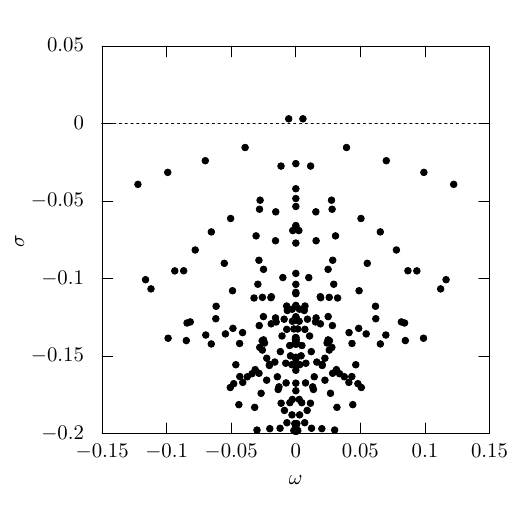}&
\includegraphics[width=0.33\textwidth]{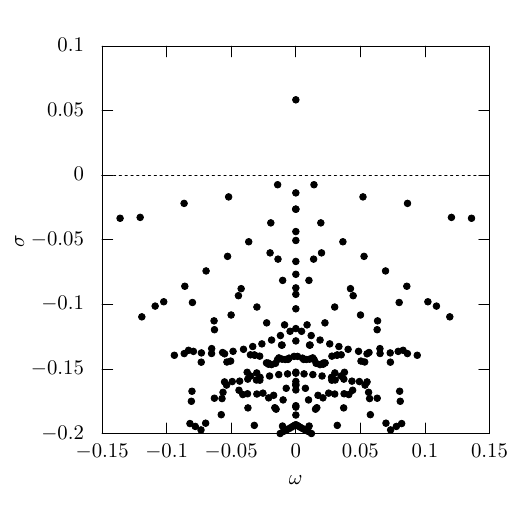}\\
(a)&(b)&(c)
\end{tabular}
\end{center}
\caption{The eigenvalues at $Re=1100$, $\beta=1$(a); $Re=1600$, $\beta=0.3$(b); $Re=1600$, $\beta=3.0$(c)}\label{a30_spectraDelta1}
\end{figure*}

For the Reynolds number $1100$, the leading unstable mode is real and is concentrated near the first recirculation bubble, as shown in Figure \ref{a30_eigenfun3D} using isosurfaces of the longitudinal vorticity component (a) and the transverse velocity component in the place of maximal amplitude (aa). The spectra for this parameters' values are presented in Figure \ref{a30_spectraDelta1}(a). Since the mode is localized in the area of the first separation bubble, it can be called a bubble mode ($B$-mode). When the transverse wavenumber $\beta$ decreases, the leading unstable eigenfunction has a noticeable, albeit smaller, amplitude in the region of the second recirculation bubble, as shown in Figure \ref{a30_eigenfun3D} (b, bb). In this case, the corresponding eigenvalue remains real. This mode couple first and second recirculation bubbles and can be named $C$-mode. In Figure \ref{a30_neutralDelta1} the stability boundary of the $B$ - mode is indicated by solid circles, and the stability boundary of the real $C$-mode is indicated by solid diamonds.

As the Reynolds number increases, there is a complex $C$ - mode, the real and imaginary parts of which are shown in the figures \ref{a30_eigenfun3D} (c, cc, d,dd) for $Re=1600$, $\beta=0.3$. As can be seen from the figure \ref{a30_spectraDelta1} (b), there is only a pair of complex eigenvalues above the line $\sigma=0$, while the real modes are stable. In contrast to long transverse perturbations (small $\beta$), short-wave perturbations are concentrated near the tail of the first recirculation bubble, as shown in Figure \ref{a30_eigenfun3D} (e,ee) at $Re=1600$, $\beta=3.0$. The  corresponding spectra are shown in Figure \ref{a30_spectraDelta1} (c). The eigenvalue of the leading mode is real. This fashion can be called a tail fashion ($T$-mode). The stability boundary of the $T$-mode is shown by solid rectangles in the figure \ref{a30_neutralDelta1}. It should be noted that a clear threshold exists only when the real $B$-mode changes the complex $C$-mode. The transition between the real $B$, $C$ and $T$ modes when the $Re$ or  $\beta$ change is carried out by  progressive increase of the velocity amplitude in one bubble and decrease in the other.

% \clearpage
% \lipsum[1-9]

\begin{figure*}[bt]
\begin{center}
\begin{tabular}{cc}
(a)&
\includegraphics[width=0.9\textwidth]{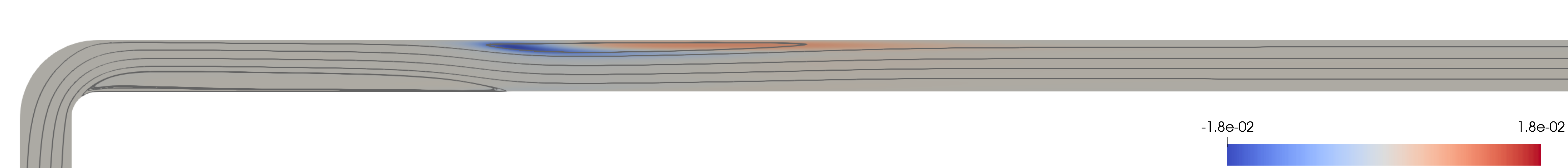}
\\
(b)
&
\includegraphics[width=0.9\textwidth]{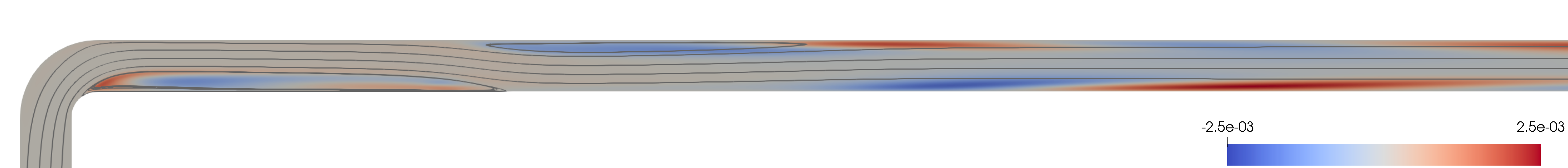}
\\
\end{tabular}
\end{center}
\caption{$v_z$ amplitude  over the base flow streamlines for selected minor modes, section at the maximal amplitude. $Re=1400$, $\beta=1.0$(a),$\beta=0.4$(b)}\label{a30_eigenfun3D_minor}
\end{figure*}

Figure \ref{a30_eigenfun3D_minor} shows selected minor eigenfunctions  at $Re=1400$: $\beta=1.0$(a), $\beta=0.4$(b,real part). In the figure \ref{a30_spectraDelta1modes} these modes are indicated by filled rectangles and empty circles, respectively. From figure \ref{a30_spectraDelta1modes} ones can see that several periodic modes have forks. As example, the leading periodic $C$-mode (empty rectangles in Figure \ref{a30_spectraDelta1modes}) at $\beta\approx0.44$ branches into two real ones (filled rectangles in the same figure), upper and lower. There is a real $C$-mode exists near the branching point. With a further increase in $\beta$, the upper branch corresponds to instability inside the recirculation bubble $V1$, and the lower branch corresponds to $V2$. The patterns shown in figure \ref{a30_eigenfun3D_minor} (b) correspond to the oblique modes that are observed in the plane channel. It can significantly affect the stability with the small curvature channel\cite{park2014streamwise}.

The dependences of the critical Reynolds numbers on the bending radius are given in table \ref{a30_critReFromDelta}. The critical Reynolds number increases almost ten times by increasing the bending parameter, and when $\delta=2.5$ it has the same order of magnitude as for the plane flow. The corresponding critical modes are real.

\begin{table}
\caption{The critical Reynolds numbers $Re_{\ast}$ and the corresponding spanwise wavenumbers $\beta_{\ast}$}
\label{a30_critReFromDelta}
\begin{center}
\begin{tabular}{|c|c|c|c|c|c|}
\hline
$\delta$                     &$0.7$     &$1.0$     &$1.5$   &$2.0$   &$2.5$\\
\hline
$Re_{\ast}$, $\pm 1$         &$469$     &$705$     &$1187$  &$1852$  &$2737$\\
\hline
$\beta_{\ast}$, $\pm 0.01$   &$1.01$    &$0.99$    &$0.96$  &$0.89$  &$0.79$\\
\hline
\end{tabular}
\end{center}
\end{table}

\section{Conclusions}\label{a30_conclusion}

Two-dimensional flow regimes in a curved channel have been investigated. It was found that as the Reynolds number increases, three recirculation bubbles appear sequentially. At relatively large Reynolds number, a regime with non-stationary vortex shedding may occur. Such vortex generation was observed earlier in a channel with a back-face step, and in a sharp bend(numerically and experimentally). The vortex shedding is being sustained due to the existence of approximation errors that produce persistent noise. The lower limit of the pulsations impossible to determine exactly. By reducing the calculation error the steady-state flow was found up to $Re \approx 1200$ in the present calculations. At $1200 < Re < 1900$, the steady-state solution was obtained using the selective frequency dumping method.  

The flow is linearly stable according to the two-dimensional perturbations at $Re<1900$, which allows to exclude the exponential growth of an unstable mode as a possible mechanism for the pulsations. It was found above that the bent flow can amplify the noise of very small amplitude $\sim 10^{-5}$. This is likely that disturbances grow up under influence of non-orthogonality effects. This scenario is supported by the results of \cite{blackburn2008convective,boujo2015sensitivity}. On the other hand, the amplitude of the resulting disturbance is large enough for noticeable influence of the nonlinear interaction mechanisms. Taking into account these considerations, it is possible to assume the existence of a two-stage scenario of subcritical transition to the pulsatile flow.

There is a range of spanwise wave numbers $\beta$ where leading three-dimensional modes grow. These modes can be periodic when the corresponding eigenvalue is complex or monotonic for real eigenvalues. The periodic mode has significant amplitude inside the first and second separation bubbles. It is the leading mode for smaller $\beta$ in the range of instability. If the spanwise wave-number $\beta$ increases, the periodic mode splits into two monotonic modes, which correspond to the first and second recirculation bubbles. The amplitude of the leading real unstable mode concentrates in the first recirculation bubble. Based on the results of \cite{barkley2002three,griffith2007wake,lanzerstorfer2012global,lupi2020global}, we can list possible mechanisms of instability occurrence in the bent channel: centrifugal\cite{barkley2002three,griffith2007wake, lanzerstorfer2012global}, elliptical \cite{griffith2007wake, lanzerstorfer2012global}, lift-up \cite{lanzerstorfer2012global}, shear instabiblity of the backflow \cite{lupi2020global}. Since the flow is complex in space, they can act simultaneously and it is difficult to detect one of them. with high degree of certainty the lift-up mechanism can be excluded, since unstable eigenfunction does not contain alternating slow and fast streaks. In the case of the dominance of shear instability, one would expect a large amplitude of the eigenfunction in the region of the largest gradient of the reverse flow, which is not confirmed by the results. Therefore, the main possible mechanisms of instability at $\delta=1$ can be identified as elliptical and centrifugal.

\section{Data Availability}
The data that support the findings of this study are available from the corresponding author upon reasonable request.

\section{*}
The manuscript is under consideration in Physics of Fluids.

\bibliographystyle{unsrt}
\bibliography{reference30}

\end{document}